\documentclass{elsart}

\usepackage{graphics}
\usepackage{graphicx}
\usepackage{epsfig}

\usepackage{amssymb}

\usepackage{slashbox}

\usepackage{amsmath}
\begin{document}

\begin{frontmatter}
\title{Linear Complexity  of Ding-Helleseth  Generalized Cyclotomic Binary Sequences of Any Order}
\thanks[label1]{This work was supported by the open fund of  State Key Laboratory of Information Security
, the Natural Science Fund of Shandong Province (No.ZR2010FM017)
, the National Natural Science Foundations of China
(No.61063041)and the Fundamental Research Funds for the Central Universities(No.10CX04038A)}
\author[label1,label2]{Tongjiang Yan} \ead{yantoji@163.com}

\address[label1]{College of Mathematics and Computational Science,
 China University of Petroleum, Dongying, 257061, China.}

\address[label2]{State Key Laboratory of Information Security (Graduate
University of Chinese Academy of Sciences), Beijing, China.}

\begin{abstract} This paper gives  the linear complexity of  binary Ding-Helleseth generalized cyclotomic sequences of  any order.
\end{abstract}

\end{frontmatter}
\newtheorem{definition}{Definition}
\newtheorem{remark}{Remark}
\newtheorem{proof}{Proof}
\section{Introduction}
Pseudo-random sequences are
widely employed in  global positioning systems, code-division multiple-access systems, radar
systems, spread-spectrum communication systems and especially in
stream ciphers. Linear complexity is
one of the important characteristics that indicates the
unpredictability of the pseudo-random sequence. If a binary pseudo-random sequence $s=(s_{0},s_{1},
s_{2},\cdots )$ satisfying
\begin{eqnarray}
s_{j}+c_{1}s_{j-1}+\cdots+c_{L}s_{j-L}=0,\,\,\,j\geq L,
\end{eqnarray}
where $L$ is a positive integer, $c_{1},c_{2},\cdots ,c_{L}\in GF(2)$,
$GF(2)$ denotes a finite Galois field of order $2$, then the least
$L$ is called the {\em linear complexity } of the sequence
$s$, denoted by $L(s)$. By Berlekamp-Massey
Algorithm, $s$ can be calculated from $2L(s)$ successive bits of itself. So, if $L(s)$ is not far less than one half of the least period of $s$
, then this sequence is thought to be good from the viewpoint
of linear complexity. Assume $S(x)=s_{0}+s_{1}x+s_{2}x^{2}+\cdots +s_{N-1}x^{N-1}$, the polynomial
\begin{eqnarray}m(x)=(x^{N}-1)/\mathrm{gcd}(S(x),x^{N}-1)\end{eqnarray}
is called the minimal generative polynomial of $s$, and we have the famous
conclusion that
\begin{eqnarray}\label{eq3}
L(s)=deg(m(x))=N-deg(\mathrm{gcd}(x^{N}-1,S(x)).
\end{eqnarray}
For more details about linear complexity , please refer \cite{Golomb,Massey}.

Cyclotomy and generalized cyclotomy are  ancient mathematic theories related many important academic fields. Since the designment of each binary pseudo-random sequence of period $N$ is essentially a partition of the residue class residue $Z_N$, then, naturally, cyclotomy and generalized cyclotomy theories play important roles in this work.

Ding-Helleseth generalized cyclotomy was proposed by C. Ding  and T. Helleseth in 1998 \cite{new}. The corresponding Ding-Helleseth  generalized cyclotomic classes were defined as the following:

Let $p$ and $q$ be two odd primes with $N=pq,2n=gcd(p-1,q-1)$,
$e=(p-1)(q-1)/2n$, $p<q$. Then we get a subgroup of the residue
ring $Z_{N}$ with its multiplication
$Z^{*}_{N}=\{g^{s}x^{l}:s=0,1,\cdots,e-1;l=0,1,\cdots,2n-1\} ,$
where $g$ is a fixed common primitive root of $p$ and $q$, and $x$
is an integer satisfying $ x\equiv g\bmod p,x\equiv 1\bmod q$.
$$D_i=\{g^{2nt+i}x^{l}:t=0, 1,
\cdots, \frac{e-2n}{2n}, l=0,1,\cdots,2n-1\}, i=0,1,\cdots,2n-1,$$ are defined {\em
Ding-Helleseth generalized cyclotomic classes} of order $2n$ with respect $p$
and $q$(D-GC).
\begin{align*}
D_i^{(p)}=\{g^{2nt+i}:t=0, 1, \ldots, \frac{p-2n-1}{2n}\} ,&
D_i^{(q)}=\{g^{2nt+i}:t=0, 1, \ldots,
\frac{q-2n-1}{2n}\},
\end{align*}
then
$gD_{i}^{(p)}=D_{i+1}^{(p)}, gD_i^{(q)}=D_{i+1}^{(q)},i=0,1,\ldots,2n-1$.

If we assume
\begin{align*}
B_m&=\bigcup_{i=m}^{m+n-1}D_{i},& B_{m}^{(p)}&=\bigcup_{i=m}^{m+n-1}D_{i}^{(p)},&B_{m}^{(q)}&=\bigcup_{i=m}^{m+n-1}D_{i}^{(q)},&&\\
P_{m}&=pB_{m}^{(q)},&P&=P_{0}\cup P_{n},&Q_{m}&=qB_{m}^{(p)},&Q&=Q_{0}\cup Q_{n},
\end{align*}
where $m=0,1,\cdots,2n-1$,then the residue class ring $Z_{pq}$ has a partition
$Z_{pq}=C_{0}\cup C_{1}$, where
\begin{align*}
C_{0}=\{0\}\cup P_{0}\cup Q_{0}\cup B_0,
C_{1}=P_{n}\cup Q_{n}\cup B_m.
\end{align*}

Ding-Helleseth generalized cyclotomic sequence
$s=(s_0,s_1,\ldots,s_i,\ldots)$ is defined as the characteristic
sequence of $C_{1}$, namely, $s_i=1$ if and only if $i\in C_1$\cite{new,d2,bainew,yanf}. This
sequence is balanced by its definition.

\section{Linear Complexity of Ding-Helleseth generalized cyclotomic sequences }

Let $\alpha$ be a primitive $N$th root of unity over the field
$\mathbf{GF}(2^L)$ that is the splitting field of $x^N-1$. Then, by
equation (\ref{eq3}), we have
\begin{align}\label{l2}
    L(s^{\infty})=N-|\{t: S(\alpha^k)=0, 0\leq k\leq N-1\}|,
\end{align}
where $S(x)$ is defined by
\begin{align}
S(x)=\sum\limits_{t\in P_{n}\cup Q_{n}\cup B_m}x^t\in\mathbf{GF}(2)[x].
\end{align}

Note that
\begin{align}\label{s1}
      S(1)&=S(\alpha^0)=\displaystyle\frac{p-1}{2}+\displaystyle\frac{q-1}{2}+\displaystyle\frac{(p-1)(q-1)}{2}\\
      &=\left  \{
    \begin{array}{ll}
    \displaystyle\frac{p-1}{2}+\displaystyle\frac{q-1}{2}\pmod 2, & \mbox{if} \,\,\,n=1. \\
    0, & \mbox{otherwise.}
    \end{array}\right.
\end{align}

The following Lemmas \ref{pqr}-\ref{pqz}, are needed to prove Lemma
\ref{zn}. They can be proved directly by the related definitions.
\newtheorem{lemma}{Lemma}
\begin{lemma}\label{pqr} \cite{bainew}
Let the symbols be the same as before.
1) If $a\in P$, then $aP=P$, $aQ=R$. \,\,\,2) If $a\in Q$, then
$aP=R$, $aQ=Q$.
\end{lemma}

\begin{lemma}\label{di} Let the symbols be the same as before.

1) If $a\in D_{i}$, then $aP_{j}=P_{i+j}, aD_j=D_{i+j}, \,\,i, j=0, 1, \cdots, 2n-1$.\,\, \\
2) If
$a\bmod p\in D_{i}^{(p)}$, then $aQ_{j}=Q_{i+j},\,\, i, j=0, 1, \cdots, 2n-1$. \\
3) If $a\bmod q\in D_{i}^{(q)}$, then $aP_{j}=P_{i+j},\,\,
i, j=0, 1, \cdots, 2n-1$.
\end{lemma}

\begin{lemma}\label{pqz} Let the symbols be the same as before.
Then
$$\sum\limits_{j\in P}\alpha^{j}=\sum\limits_{j\in Q}\alpha^{j}=\sum\limits_{i\in
Z_{N}^*}\alpha^{j}=1.$$
\end{lemma}

\begin{lemma}\label{zn} If  $k_1,k_2\in D_i$, and $k_1\bmod p\in D_{j}^{(p)},k_2\bmod p\in D_{j+n}^{(p)}$ for any $i$ and
$j$, then $S(\alpha^{k_1})+S(\alpha^{k_2})=1$.
\end{lemma}
\noindent\textbf{Proof.} If $k_1,k_2\in D_i$, and $k_1\bmod p\in D_{j}^{(p)},k_2\bmod p\in D_{j+n}^{(p)}$ for any $i$, by Lemmas \ref{pqr} and \ref{di}, then
$$\begin{array}{llll}
  S(\alpha^{k_1})+S(\alpha^{k_2})&=&\sum\limits_{t\in P_{n}\cup Q_{n}\cup B_n}\alpha^{k_1t}+\sum\limits_{t\in P_{n}\cup Q_{n}\cup B_n}\alpha^{k_2t}\\
  &=&\sum\limits_{t\in k_1P_{n}\cup k_1 Q_{n}\cup k_1B_n}\alpha^{t}+\sum\limits_{t\in k_2P_{n}\cup k_2Q_{n}\cup k_2B_n}\alpha^{t}\\
  &=&\sum\limits_{t\in P_{n+i}\cup Q_{n+j}\cup B_{n+i}}\alpha^{t}+\sum\limits_{t\in P_{n+i}\cup Q_{j}\cup B_{n+i}}\alpha^{t}\\
  &=&\sum\limits_{t\in Q_{n+j}\cup Q_{j}} \alpha^{t}=\sum\limits_{t\in Q} \alpha^{t}=1.
\end{array}$$
From the above Lemma \ref{zn} and the symmetry of $Z_N^{*}$, we have
\begin{lemma}\label{bspq} Let the symbols be the same as before. Then
$$
|\{t: S(\alpha^k)=0,  k\in Z_N^{*}\}|\leq \frac{(p-1)(q-1)}{2}
$$
\end{lemma}
The following, Lemma \ref{pq}, is needed to prove Lemma \ref{spq}:

\begin{lemma}\label{pq} Let the symbols be the same as before. Then
\[ \sum\limits_{i\in D_j}\alpha^{ki}= \left\{
\begin{array}{llll}
 0, & \mbox{if $k\in P$},\\
\displaystyle{\frac{q-1}{2n}}\pmod{2}, & \mbox{if $k\in Q$,}
\end{array}
\right.\,\,j=0, 1, \cdots, 2n-1.
\]
\end{lemma}
\noindent\textbf{Proof.} Similar to the proof of Lemma 2 in
\cite{bainew}.

\begin{lemma}\label{spq} Let the symbols be the same as before. Then
$$
S(\alpha^{k})= \left\{
\begin{array}{llll}
\displaystyle\frac{p-1}{2}+\sum\limits_{t\in P_n}\alpha^{kt}\pmod{2},& \mbox{if $k\in P$ },\\
\sum\limits_{t\in Q_n}\alpha^{kt}\!\!\!\!\pmod{2},  & \mbox{if $k\in Q$}.\\
\end{array}
\right.
$$
\end{lemma}

\noindent\textbf{Proof.} If $k\in P$, then, by Lemma \ref{pqr} ,
$kQ=\{0\}$; by Lemma \ref{pq}, $\sum\limits_{t\in
D_i}\alpha^{kt}=0$. It follows
that
$$\begin{array}{llll}
  S(\alpha^{k})&=&\sum\limits_{t\in P_{n}\cup Q_{n}\cup B_n}\alpha^{k_1t} \\
 &=&\sum\limits_{i\in P_n}\alpha^{ki}+\displaystyle\frac{p-1}{2}\pmod{2}.\\
\end{array}$$

If $k\in Q$, then, by Lemma \ref{pqr}, $kP=\{0\}$, and by Lemma
\ref{pq}, $\sum\limits_{t\in D_i}\alpha^{kt}=\displaystyle\frac{q-1}{2n}\pmod{2}$. It
follows that
$$\begin{array}{llllll}
  S(\alpha^{k})&=&\sum\limits_{t\in P_{n}\cup Q_{n}\cup B_n}\alpha^{kt} \\
  &=&\sum\limits_{t\in Q_{n}}\alpha^{kt}+\displaystyle\frac{q-1}{2}
\end{array}$$
This lemma is proven.

Assume $T_a^{(p)}(\alpha)=\sum\limits_{t\in P_a}\alpha^t, T_b^{(q)}(\alpha)=\sum\limits_{t\in Q_b}\alpha^t, a,b=0,1,\cdots,2n-1$. Similarly to Lemma \ref{zn}, the following Lemma \ref{st} can be given.
\begin{lemma}\label{st} Let the symbols be the same as before. Then
$$
|\{t: T_a^{(p)}(\alpha^k)=0,  k\in P\}|\leq \frac{q-1}{2}, |\{t: T_a^{(p)}(\alpha^k)=0,  k\in Q\}|\leq \frac{p-1}{2}.
$$
\end{lemma}
\noindent\textbf{Proof.}
$T_a^{(p)}(\alpha^k)+T_{a+n}^{(p)}(\alpha^k)=\sum\limits_{t\in P_a}\alpha^t+\sum\limits_{t\in P_{a+n}}\alpha^t=1.$
By symmetry, we have $|\{t: T_a^{(p)}(\alpha^k)=0,  k\in P\}|\leq \frac{q-1}{2}$.
Similarly, $T_b^{(p)}(\alpha^k)+T_{b+n}^{(p)}(\alpha^k)=1$, then  $|\{t: T_a^{(p)}(\alpha^k)=0,  k\in Q\}|\leq \frac{p-1}{2}.$

Based on Equation \ref{s1} and Lemmas \ref{bspq},\ref{spq},\ref{st}, The following  Theorem  \ref{p5q5} can be obtained directly.

\newtheorem{theorem}{Theorem}
\begin{theorem}\label{p5q5}
Linear complexity of  Ding generalized cyclotomic sequences for any order
$$L(s^\infty)\geq \frac{pq-1}{2}.$$
\end{theorem}

\begin{remark} In \cite{yanf}, a mistake has been found in Table 1 in Lemma 4 by Yuan Sun in 2008, which leads to a mistake in the final result.
\end{remark}

\end{document}